\documentstyle[11pt,aspconf]{article}

\begin{document}

\title{The HST Spectrum of I Zw 1: Implications of the 
C~III$^*~\lambda 1176$ Emission Line}

\author{Ari Laor}
\affil{Physics Department, Technion, Haifa 32000, Israel}

\author{Buell T. Jannuzi; Richard F. Green and Todd A. Boroson}
\affil{National Optical Astronomy Observatories, Tucson, AZ 85719}

% \altaffiltext{1}{Visiting Astronomer, Cerro Tololo Inter-American Observatory. 

\begin{abstract}
I Zw 1 is a well known narrow line quasar 
with very strong Fe~II emission. High S/N spectra  
obtained with the HST {\em FOS} 
show a remarkably rich emission line spectrum. 
The C~III$^*~\lambda 1176$ line is clearly detected in emission for
the first time in AGNs. This line
arises from radiative decay to the $2s2p^3P^o_{0,1,2}$ metastable levels
of C~III. The observed flux is {\em $\sim 50$ larger} 
than expected from
collisional excitation, or dielectronic recombination, in photoionized
gas. The most plausible mechanism for the large
enhancement in the C~III$^*~\lambda 1176$ flux is resonance scattering of
continuum photons by C~III$^*$ ions. This mechanism requires 
{\em large velocity gradients} ($\sim 1000$~km~s$^{-1}$) within 
each emitting cloud
in the BLR. Such large velocity gradients can be induced by forces 
external
to the gas in the BLR clouds, such as tidal disruption, or 
radiation pressure.
\end{abstract}


\section{Introduction}

A very weak feature at 1176~\AA\ was found by Laor et al. (1995) 
in three high S/N quasar spectra obtained by HST, and was tentatively
identified as C~III$^*~\lambda 1176$. The rich UV emission line spectrum
of the narrow line quasar I Zw 1 obtained by HST (Laor et al. in 
preparation) allowed us 
to clearly identify this feature as C~III$^*~\lambda 1176$. This line
was previously detected in AGNs only in absorption (Bromage et al. 1985,
Kriss et al. 1992). The line originates from radiative transition of
electrons at the $2p^2\ ^3P_{0,1,2}$ levels (17.1 eV above the ground 
level), down to the $2s2p^3P^o_{0,1,2}$ levels (6.5 eV above the ground 
level). Given the high energy of the $2p^2\ ^3P_{0,1,2}$ levels, the 
presence of significant C~III$^*~\lambda 1176$ emission appears surprising.

\section{The Calculations}

In order to calculate the 
C~III$^*~\lambda 1176$ line flux we solved the equilibrium
equations for the population of the lowest 10 levels of C~III (n=2
levels). Collisional coupling of all levels, and radiative decay
to all levels are included. The contributions of recombination 
(mostly dielectronic) and 
continuum fluorescence to the level population were not included 
since they are not important for the observed line flux.

We find that for typical BLR conditions which are able to generate 
significant C~III]$~\lambda 1909$ emission ($n_e\le 10^{10}$), 
the $f(1176)/f(1909)$ ratio is
at least 50 times smaller than observed.

\section{Why is C~III$^*~\lambda 1176$ strongly enhanced?}

\subsection{A High Density Component in the BLR?}
The observed $f(1176)/f(1909)$ ratio is obtained for $\log n_e=11.5$. But 
this ratio is obtained because $\lambda 1909$ is strongly suppressed 
at such a high density, and not because $\lambda 1176$ is enhanced. 
A high density component cannot produce the observed $\lambda 1176$ 
flux even for a covering factor of $\sim 100$\% 

\subsection{Dielectronic Recombination?}

Dielectronic recombination is ruled out based on line ratios. 
It predicts a ratio of 2.5 for $f(2297)/f(1176)$, compared with an 
observed ratio $<0.2$.

\subsection{Collisionally Ionized Gas?}
The observed $f(1176)/f(1909)$ is obtained for 
$T>2.5\times 10^4$~K. Such a component was also inferred by Kriss et al. 
(1992) in NGC~1068, based on the C~III$~\lambda 977$ and N~III$~\lambda 990$ 
lines. However, the overall observed emission line spectrum is fit well by 
photoionization models, rather than collisional ionization models.

\subsection{Resonance Scattering?}
The $EW$ produced by resonance scattering of continuum photons 
in the BLR is:  
$ EW=C\lambda \Delta v/c\times {\rm min}(1,\tau) $,
where $C$ is the BLR covering factor, $\Delta v$ is the
velocity dispersion within the cloud, and $\tau$ is the line center 
optical depth (for $\tau\gg 1$, the $EW$ is increased by 
$\sqrt{\ln \tau}$). The optical depth is
$ \tau=1.5\times 10^6 \Delta v_{10}^{-1} \Sigma_{i,j} n_if_{ij} $,
where $\Delta v=10\Delta v_{10}$~km~s$^{-1}$, and the sum is over
the 6 permitted transitions contributing to the $\lambda 1176$ line.
We get $ \tau=1000$-$6000\Delta v_{10}^{-1}$ for $\log n_e=$8-10. 
Thus the observed 
C~III$^*~\lambda 1176$ $EW$ of 1.4~\AA\ can be produced if
$\Delta v={\rm FWHM}(\lambda 1176)=1000$~km~s$^{-1}$, 
and $C\sim 0.35$. A similar process of
continuum fluorescence was invoked by Ferguson et al. (1995) to
explain the strong C~III$~\lambda 977$ and N~III$~\lambda 990$ 
lines in NGC~1068.

\acknowledgments

Support for this work was provided by NASA through grant number 
GO-5486.01-93A from the Space Telescope Science Institute.

\end{document}